# W+W->Z+Z Scattering at the LHC

Dan Green
(dgreen@fnal.gov)
US CMS Dept.
Fermilab

The process of vector boson scattering is of great interest because it is completely prescribed in the Standard Model (SM) once the Higgs mass is known. The LHC is a source of Z-Z pairs produced by means of the vector boson fusion mechanism which is inferred by the presence of two additional forward (or tag) jets. Because the reaction can be initiated by valence quarks and because the W+W->Z+Z cross section approaches a constant in the SM, the scattering process is accessible at the LHC.

## The Fundamental Process, W+W->Z+Z

The cross section for the process W+W -> Z+Z at high C.M. energies depends only on the vacuum expectation value of the Higgs field (or the W mass) and the electroweak coupling constant, $\alpha_W$. The three Feynman diagrams contributing to this fundamental process are shown in Fig. 1.

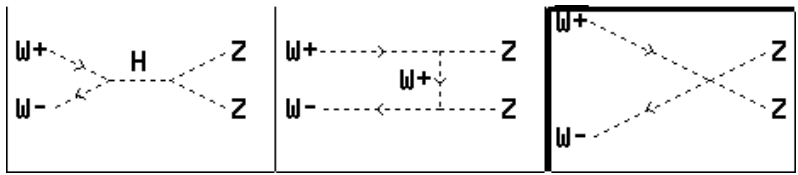

Figure 1: Feynman diagrams for W+W->Z+Z scattering in the Standard Model

These three diagrams correspond to s channel virtual Higgs production and decay, t channel W exchange, and a quartic WWZZ coupling. The triple boson coupling has been measured at LEPII [1] and will be measured at the Tevatron before LHC data taking begins. Dimensional analysis of the Feynman diagrams allows on to estimate the cross section at asymptotic energies.

$$\hat{\sigma}(W+W \to Z+Z) \sim \alpha_W^2 / M_W^2 = 68 \; pb \tag{1}$$

The angular distribution can be evaluated using COMPHEP (4.2.0) [2]. The result is shown in Fig. 2. Note that the distribution is very forward, backward peaked. This peaking reflects the cancellations between the three Feynman diagrams, which are, by themselves, rather isotropic but divergent with Z-Z C.M. energy.

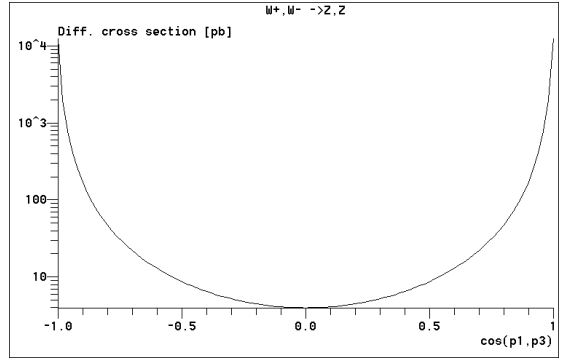

Figure 2: The distribution of the cosine of the angle of the Z with respect to the incident W in the W-W C.M. frame.

The cross section for the process W+W->Z+Z is displayed in Fig.3 as a function of W-W C.M. energy. In this plot a Higgs mass of 130 GeV is assumed. At energies well above the Higgs resonance the cross section rises smoothly from about 300 pb. at C.M. energy of 400 GeV to ~ 328 pb. at very high energies. This value is consistent with the estimate made in Eq.1. In comparison the reversed reaction Z+Z->W+W has an asymptotic cross section of about 656 pb.

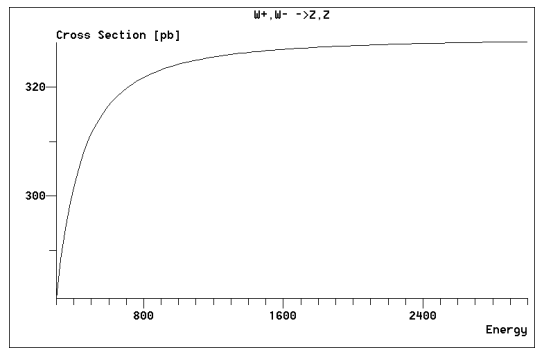

Figure 3: Cross section for the process W + W -> Z + Z as a function of W-W C.M. energy. The Higgs mass has been assumed to be 200 GeV.

## The Effective W Approximation

In order to make a simple first analytic estimate of the rate for the fundamental process at hadron colliders the effective W approximation has first been adopted. This approximation consists of perturbatively calculating the probability for a quark in the proton to emit a W boson and treating the quark as effectively a source of W bosons. The estimate for the W parton distribution function for a transverse (T) W, carrying a momentum fraction x of the quark, emitted by a quark (q) is [3] to lowest order in the electroweak fine structure constant, $\alpha_W$:

$$f_{q/W_T}(x) = (\alpha_W/8\pi)\ln(\hat{s}/M_W^2)[1+(1-x)^2]/x \tag{2}$$

The longitudinal quarks have a smaller probability. The q-q C.M. energy is $\hat{s}$. The logarithmic factor is similar to that which arises in calculating the two-photon higher order process in QED [3].

The effective luminosity, L, for the q-q system to emit a transversely polarized W-W pair can be calculated analytically using the distribution function quoted in Eq.2 [3].

$$(dL/da)_{qq/W_T W_T} = \int_a^1 f_{q/W}(x) f_{q/W}(a/x)(dx/x)$$
$$= (\alpha_W/8\pi)^2 (1/a)[\ln(\hat{s}/M_W^2)]^2 [(2+a)^2 \ln(1/a) - 2(1-a)(3+a)] \tag{3}$$

The differential luminosity in the q-q system, dL/da, comes from convoluting the W distribution functions over all allowed momentum fractions of the emitted W bosons. The luminosity is a function only of $a = M^2/\hat{s}$, where M is the W-W invariant mass.

In the case of longitudinal polarization the differential luminosity can also be done in closed form [4].

$$(dL/da)_{qq/W_L W_L} = \int_a^1 f_{q/W}(x) f_{q/W}(a/x)(dx/x)$$
$$= (\alpha_W/4\pi)^2 (1/a)[(1+a)\ln(1/a) - 2(1-a)] \tag{4}$$

The final step is to find the cross section for p-p scattering using the distribution functions, $f_1(x)$ and $f_2(x)$, for the quarks themselves. The luminosity for W-W scattering in p-p scattering is estimated by convoluting the q-q luminosity for W-W over all possible quark momentum fractions as illustrated in Eq.5.

$$(dL/d\tau)_{pp/W_T W_T} = \int_\tau^1 dp/p \int_p^1 (dx/x) f_1(x) f_2(p/x)(dL/da)_{qq/W_T W_T} \tag{5}$$
$$\tau = M^2/s, \, p = \hat{s}/s, \, a = \tau/p$$

The p-p C.M. energy squared is s. The p-p cross-section as a function of W-W mass, is then the product of the differential luminosity and the fundamental W-W cross-section at mass M.

$$d\sigma_{pp/WW}/d\tau = (dL/d\tau)_{pp/WW} \sigma_{WW\to ZZ}(M^2 = \tau s) \tag{6}$$

This expression is not obtainable in closed analytic form. However, numerical integration is reasonably straightforward, allowing for an evaluation of the cross section at the LHC in this approximation. To set the scale, an order of magnitude estimate of, $(\alpha_W/4\pi)^2[\sigma = 330 pb] = 2.3$ fb., gives the right order of magnitude for the p-p cross section at the LHC with longitudinal polarization.

Note that previous studies of Higgs production via vector boson fusion have shown that this process is a very competitive way to search for a Standard Model (SM) Higgs [5]. In fact, at high Higgs mass the cross section for vector boson fusion production of a SM Higgs is a large fraction of the total Higgs production cross section [6]. In a sense this study generalizes that work from resonant vector boson scattering (e.g. Fig. 1) to boson-boson scattering n general.

## Z+Z+J+J Production in p-p Scattering - Numerical

Since the quark distribution functions which were used in this note [2] are well measured valence quark distributions, the error associated with them is small. Note that for x > 0.3 the valence distribution for the u quark is an order of magnitude larger than the sea distribution function. Hence at high Z-Z mass valence processes should dominate the cross section.

The result of the numerical integration for transverse W polarization in Eq.6 is shown in Fig 4. Both valence-sea and valence-valence quark-quark convolutions are shown. Clearly, sea-sea would be even smaller. The valence-valence process dominates, as expected

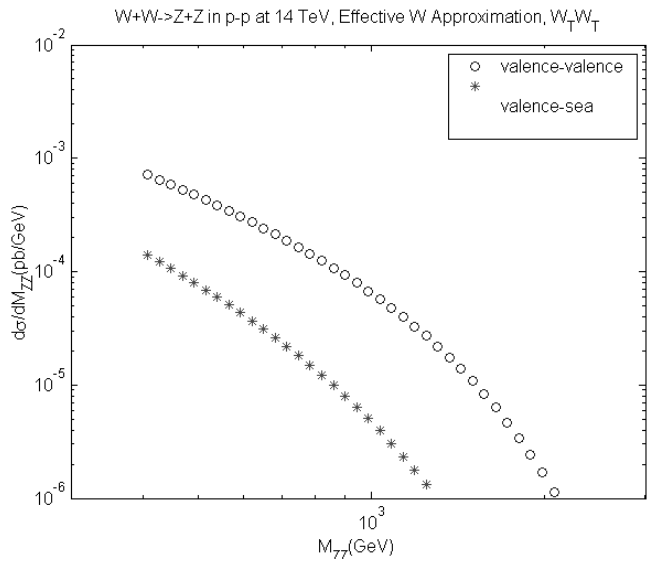

Figure 4: Differential cross section in p-p scattering at the LHC for the vector boson fusion production of Z-Z pairs as a function of the Z-Z mass.

The cross section at the LHC (14 TeV) for Z-Z masses greater than 400 GeV, in the effective W approximation and assuming a $W_T W_T$ initial state, is ~ 0.17 pb., a factor ~ 2000 less than the cross section for the fundamental W-W process. The valence-sea cross section is 0.023 pb., a factor ~8 less than the valence-valence value. On the basis of these estimates, the sea contributions to this process can be dropped. The numerical values agree with other estimates of W-W fusion [7].

There are other contributions to Z+Z+J+J production. The fundamental process, Z+Z->Z+Z, proceeds by Feynman diagrams corresponding to s channel Higgs decay and by t channel Higgs exchange. The cross section is ~ 2.0 pb. at a Z-Z mass of 1 TeV, falling to ~ 0.4 pb. at 2 TeV mass. As seen in Fig. 3, the W+W->Z+Z process is ~ 200 times larger, so that Z+Z production need not be considered further.

Note also that the effective W approximation assumes that the W-W mass M is much greater than the W mass. It is for that reason the W-W mass M is restricted to be above 400 GeV. The approximation also assumes that the parameter a is much less than one. In Fig. 4 the numerical evaluation was done with a < 0.1. If the evaluation is done with a < 0.2 or a < 0.05, the cross section above 400 GeV Z-Z mass varies by a factor about two. Obviously, the estimate for the cross section is not very stable.

The p-p energy dependence of the vector boson fusion process is quite pronounced. In Fig. 5 is shown the cross section as a function of Z-Z mass at a p-p C.M. energy of 2 and 14 TeV. At a Z-Z mass of 400GeV the cross section at the LHC is a factor ~ 20000 larger than at the Tevatron and that factor increases rapidly with Z-Z mass. Therefore, the study of this process at the LHC is much easier than at the Tevatron. In what follows, only the LHC energy of 14 TeV will be considered.

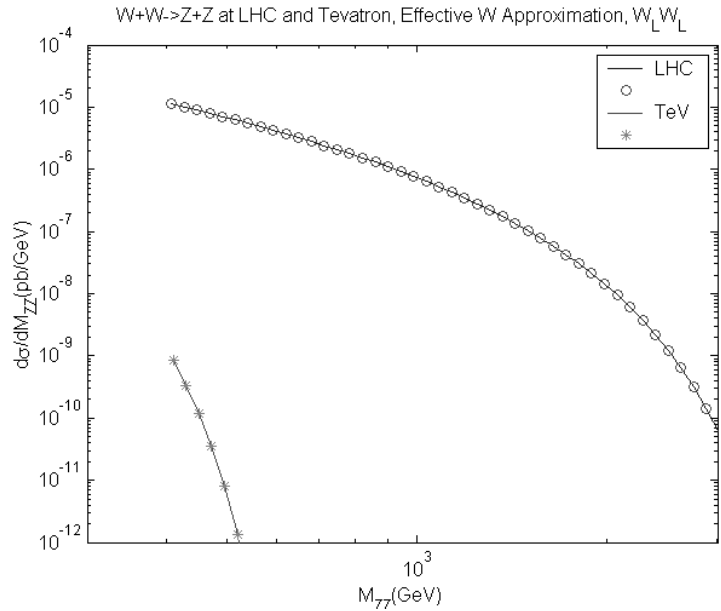

Figure 5: Cross section for p-p production of Z-Z pairs by means of the vector boson fusion mechanism at a p-p C.M. energy of 2 and 14 TeV.

The differential cross section as a function of Z-Z mass is shown in Fig.6 in the case of both transverse and longitudinal initial WW states. Note that the cross section difference is a factor ~ 80 between the two cases.

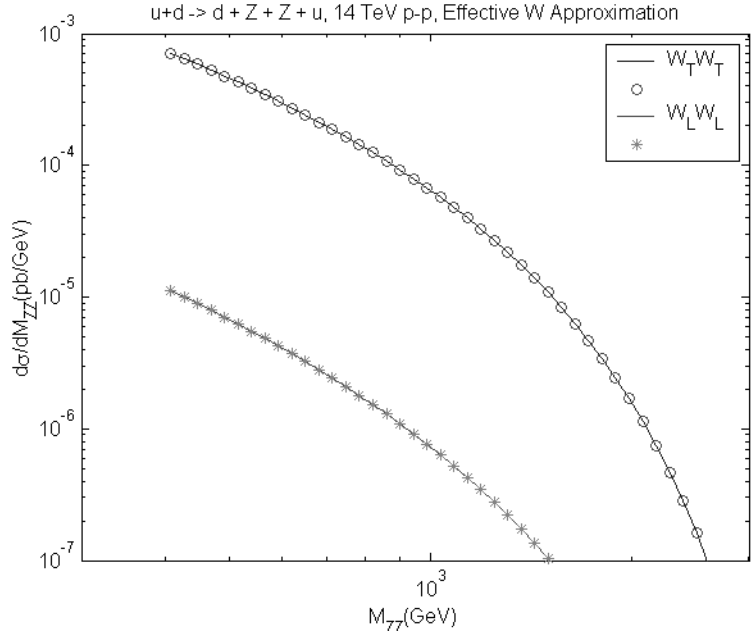

Figure 6: Differential cross section as a function of Z-Z mass in the effective W approximation in 14 TeV p-p interactions for the fundamental process W+W -> Z + Z coupled to transverse or longitudinal W.

## Z+Z+J+J - COMPHEP Results

The COMPHEP [2] program was used to go beyond the effective W approximation. Other, earlier, studies [8] have also used exact matrix element calculations. This step is needed because the effective W approximation does not allow for a detailed study of trigger and reconstruction efficiency or geometric acceptance of the four final state particles.

The possible Feynman diagrams for the valence – valence process, u+d -> d+Z+Z+u, number fifty in all. This makes for a very complex reaction amplitude. For this reason only three Feynman diagrams were selected, as indicated in Fig.7. These "electroweak" diagrams correspond to those shown in Fig.1 for the fundamental process. The W emission processes, $u \to d + W^+$ and $d \to u + W^-$ imply that the valence state, u-d, can form the $W^+W^-$ two vector boson initial state along with the two "tag" jets, d, u which move along the forward and backward directions, roughly collinear with the proton beams. These jets "tag" the presence of the emission of the two vector bosons.

There are two other electroweak diagrams with u+d->u+Z+Z+d where the fundamental process is Z+Z->Z+Z, due to u->u+Z and d->d+Z vertices. The two diagrams are Z-Z formation of Higgs with virtual ZZ decay and t channel Z exchange. These are ignored because the cross section is small with respect to that for W+W->Z+Z as mentioned above.

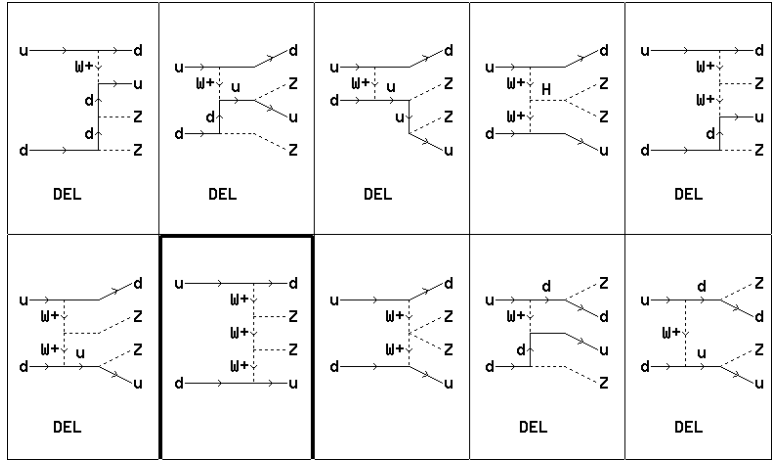

Figure 7: Ten Feynman diagrams of the fifty created by COMPHEP for the process u+d->d+Z+Z+u. Only three remain undeleted in this study. They correspond to the three electroweak diagrams contributing to the fundamental process, W+W->Z+Z, shown in Fig.1.

The generated events were analyzed assuming "2 -> 4" kinematics with double virtual W emission. Conservation of energy-momentum assuming no initial state transverse momentum requires that the initial state quarks have momentum fractions $x_1$ and $x_2$, where the four parton final state has momentum fraction x and mass squared of $\hat{s}$.

$$x_1 x_2 = \hat{s}/s,\ x_1 - x_2 = x \tag{7}$$

These two equations can be solved for the initial quark momentum fractions since the four final state partons are measured in the detectors. It remains only to properly associate the two "tag" jets with the initial quarks in order to find the two assumed W momentum vectors. In the COMPHEP analysis the choice made was to pick the association of the "tag" jet to initial quark with the lowest sum of virtual W masses.

The cross section for p- p scattering at 14 TeV C.M. energy implied by these three diagrams is shown in Fig.8 as a function of the Z-Z pair mass. The cross section integrated over all Z-Z pair masses is estimated to be 8 fb. Also shown in Fig.8 is the spectrum for a 1 TeV Higgs mass, which has a slightly larger cross section of 14 fb. The differential cross section shown in Fig.8 is in reasonable agreement with the effective W estimate for longitudinal W shown in Fig.6. The Z-Z pair mass has a mean of about 680 GeV.

The effective W approximation ignores any transverse momentum for the W bosons, while COMPHEP treats the W transverse momentum explicitly. Indeed the W transverse momentum is not small as seen in Fig.9. The outgoing u "tag" jet and the emitted virtual

W have the same transverse momentum since the initial state partons are assumed to have no transverse momenta. The virtual W mass and transverse momentum are highly correlated. The mean W mass is, $<M_W> = 210$ GeV, while the mean transverse momentum is, $<P_{TW}> = 148$ GeV. Therefore, the transverse momentum of the tag jet yields a good initial estimate for the virtual W mass.

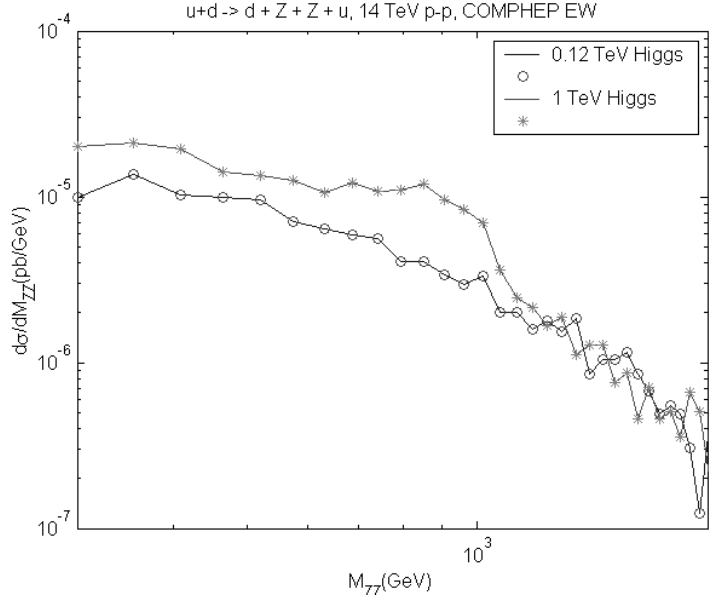

Figure 8: Cross section evaluated by COMPHEP as a function of the Z-Z pair mass for the active Feynman diagrams shown in Fig.7. The total cross section for a light Higgs mass is ~ 8 fb.

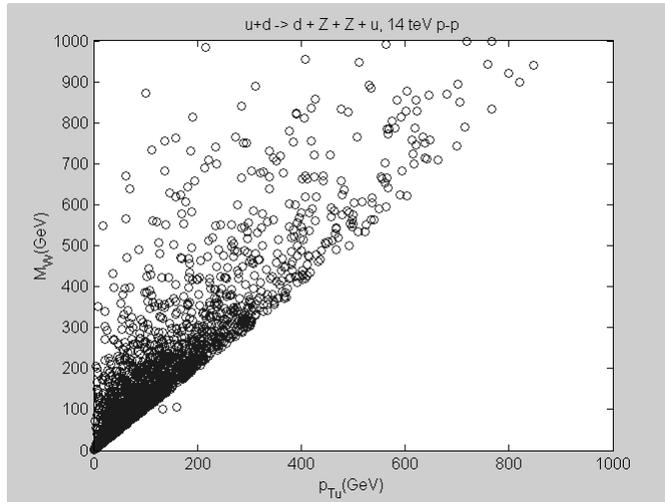

Figure 9: Transverse momentum of the final state u quark ("tag" jet) in the process, p(u)+p(d)-> d+Z+Z+u vs. the virtual W effective mass.

The average parton x values are, $<x_u> = 0.22$, $<x_d> = 0.17$. The final state "tag" jets have average transverse momentum of ~ 145 GeV. The produced Z have average transverse momenta of ~ 190 GeV. The Z are produced centrally. Almost all Z have a pseudorapidity with magnitude less than 1.5. There is a kinematic correlation between the Z-Z mass and the Z rapidities. In Fig.10 is shown the scatter plot of the sum of the Z pseudorapidities and the Z-Z mass. Higher masses are clearly more centrally produced and thus more efficiently detected.

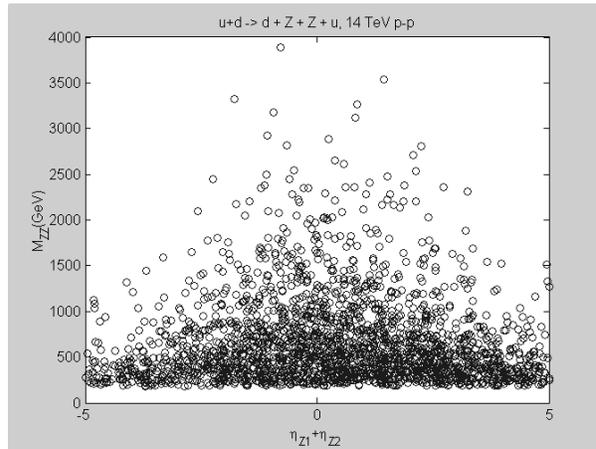

Figure 10: Scatter plot of the sum of the pseudorapidities of the two Z's and the ZZ mass. Heavier Z-Z masses are kinematically forced toward wide angles.

The "tag" jets in light Higgs production are produced at rather forward angles [4]. The same holds true here, although the mass scales are somewhat higher. A scatter plot of the pseudorapidity of the two "tag" jets is shown in Fig. 11. Clearly, they are quite well separated into different hemispheres, making triggering quite easy. The mean pseudorapidity for the u "tag" jet is 2.31, while the mean for the d "tag" jet is –2.00.

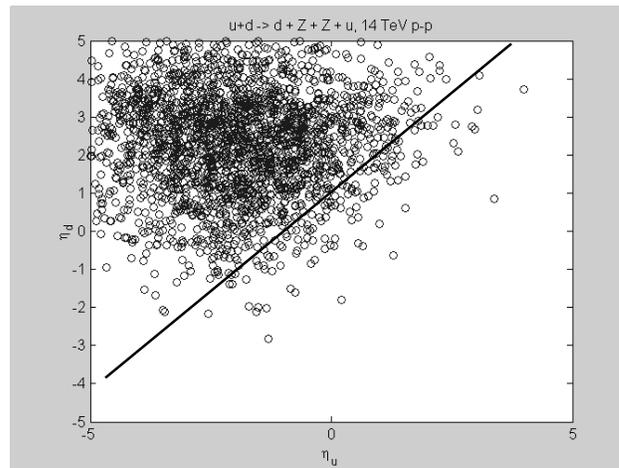

Figure 11: Scatter plot of the pseudorapidity of the two "tag" jets created by COMPHEP. Note the fairly good separation into opposite hemispheres. The line indicates the boundary where the η difference of the two "tag" jets is 1.

The COMPHEP program was also used to select only the quartic diagram. This is obviously not correct, but it serves as an indication of what might occur should the Standard Model (SM) not prove to be valid and should unitarity be imposed by some different mechanism. The cross section is about 160 fb., about 20 times larger than the SM cross section for W+W -> Z + Z. The large increase occurs because the SM result is due to large cancellations among diagrams.

The reconstructed W-W initial state is used to compute the scattering angle, which is defined to be the same as in Fig.2, the angle between the W and Z in the W-W C.M. frame. In Fig.12 the distribution of that angle is shown in the case (un-normalized) of the SM and the quartic diagram alone. Clearly, deviations from the SM might have profound influence on both the magnitude of the cross section and on the angular distribution. Note the strong forward – backward peaking in the SM case, just as in Fig.2. This result shows that, in the COMPHEP context, the fundamental W+W -> Z +Z process can be well reconstructed from the u+d -> d+Z+Z+u four body final state process in p-p interactions.

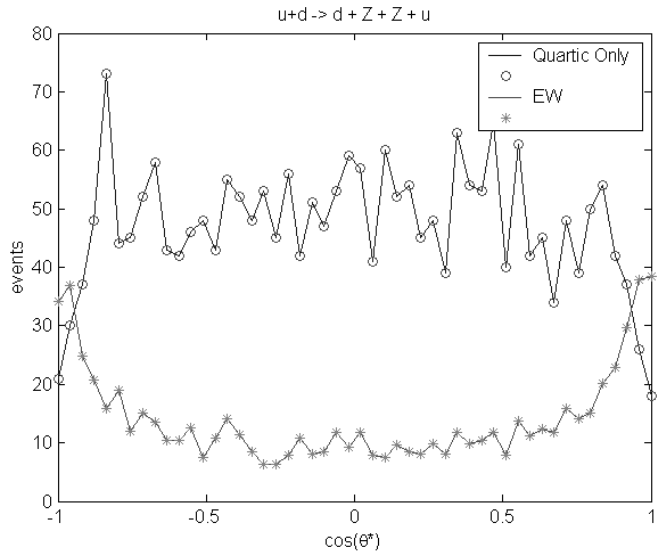

Figure 12: Angular distribution in W+W -> Z+Z scattering as reconstructed from the four body final state, d+Z+Z+u. The quartic distribution is roughly isotropic, while the SM distribution is very forward backward peaked.

## PYTHIA Results, Z+J+J+J+J

Assuming the SM is correct, the cross section due to the three electroweak diagrams given in Fig.7 above for Z+Z+J+J is ~ 16 fb. (u+d and d+u). For a one-year exposure at the LHC at design luminosity, there are only about 150 events produced where one Z decays into electrons or muons and the second decays into two jets. This process would clearly benefit from the imagined ten-fold increase in luminosity or the Super LHV (SLHC) [9]. The geometric efficiency for the leptons to occur within $|\eta| < 2.5$ and the jets ("tag" and Z decays) to have $|\eta| < 5.0$ is expected to be quite large. The mean transverse momenta are also large, making trigger cuts for jets and leptons also efficient.

Since the topology to be considered is a Z decay into dileptons plus four jets, other fundamental processes also contribute. One such process is W + Z -> W + Z. There are four contributing Feynman diagrams; quartic, t channel Higgs exchange, t channel W exchange, and W* decay into W + Z. In the Feynman gauge the W exchange diagram is the dominant one. This dominance results in a strong backward peak in the angle between the incoming and outgoing W as seen in Fig.13.

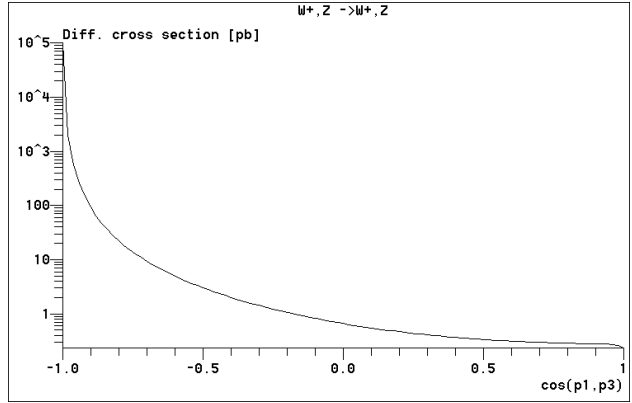

Figure 13: Angular distribution for the fundamental process W + Z -> W + Z, where the angle is between the incoming and outgoing W in the C.M. frame.

The cross section for the fundamental process is, $\hat{\sigma}(W^+ + Z -> W^+ + Z) = 310\, pb$, comparable to the cross section for W+W ->Z+Z scattering. For p-p scattering there are both $W^+$+Z and $W^-$+Z final states. In the case of $W^+$+Z, the W is emitted by $u \to d + W^+$, while the Z can be radiated by either of two valence quarks, $u \to u + Z, d \to d + Z$. The p-p cross section is ~ 2(30+12) = 84 fb., rather larger than Z+Z cross section.

Since the W and Z are not well resolved by calorimetry at the LHC, both W+Z and Z+Z processes will be in the Z+ four jet data set. If the number of events permits, leptonic decays of the W ( 22 % branching ratio) can be used in a sample with a Z leptonic decay plus a third lepton and missing energy along with the two "tag" jets. That sample would allow an independent estimate of the W+Z cross section.

In order to explore in more detail the geometric efficiency, the trigger efficiency, and the reconstruction method, the COMPHEP Z bosons should decay into leptons and quark jets and the jets should then fragment. In order to accomplish this, the program PYTHIA was used with input events from COMPHEP. The mean number of stable final state particles after decay and fragmentation was 307.

The leptons were identified and events were selected where one Z boson in the final state decayed into a muon or electron. The geometric and trigger efficiency was estimated by requiring each lepton to have a pseudorapidity magnitude < 2.5 and a transverse momentum > 10 GeV. The efficiency for this sample was ~ 74 %.

Jets were formed from the final state particles using a fixed cone algorithm. The "seed" transverse energy was taken to be 3 GeV. Because the Z have a fairly large momentum the two jets from quark fragmentation have a small opening angle. In order to then resolve the two jets, a small jet cone radius was chosen, in this case 0.3. The mass of the dijet system was calculated for the four largest transverse momentum jets (six pairs per event). The resulting distribution is shown in Fig.14.

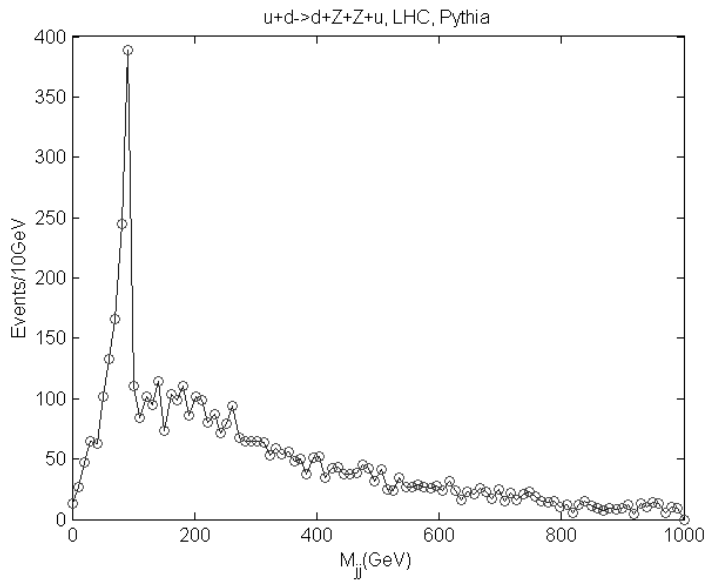

Figure 14: Dijet mass distribution for the six pair combinations contained in the four highest transverse momentum jets in the event.

A clear Z mass peak is seen. There was no attempt made to simulate detector energy resolutions. The width of the Z peak arises from the narrow cone radius which was needed to resolve the decay jets. Clearly, the reconstruction algorithm is not yet optimal. . Nevertheless, the efficiency to find the two jets and to compute the correct Z mass within 30 GeV is about 81%.  The geometric and trigger efficiencies are estimated by requiring each jet to have a pseudorapidity magnitude < 5 and a transverse momentum > 20 GeV. The dijet efficiency is about 90 %.

The efficiency to find the two "tag" jets is estimated to be about 87%. The overall efficiency to detect, trigger, and reconstruct the two lepton plus four jet final state is about 41%. Therefore, a total of about 62 Z-Z events are expected for one year of LHC running at design luminosity. There will be a larger sample, about 310 events, of W-Z events in the two lepton plus four jet final state.

Other studies [10, 11, 12, 13, 14] of the vector boson fusion process have concentrated on measuring deviations from SM couplings. This paper concentrates on more experimental questions. In order to go further than this study a complete detector simulation should be made. That work would yield a firmer estimate for the event rates and the trigger and reconstruction strategy. Also, the backgrounds to the electroweak processes should be evaluated. For example QCD radiation accompanying Z+Z production is a clear background process that needs to be studied. More detailed work is clearly called for.

## Acknowledgments

The help of Dr. H. Pi with the COMPHEP generation of events is gratefully acknowledged. Dr. H. Wenzel provided extremely useful Linux advice and ran PYTHIA on the COMPHEP events for this study.

## References

1. LEPEWWG/TGC/2002-01 (2002)
   OPAL, CERN-EP/2003-042, July 14, 2003

2. A. Pukhov et al., User's Manual, COMPHEP V33, Preprint INP-MSU 98-41/542

3. V. Barger and R. Phillips, Collider Physics, Addison-Wesley Publishing Co. (1987)

4. J. Gunion, H. Haber, G. Kane, S. Dawson, The Higgs Hunter's Guide, Addison-Wesley Publishing Co., (1990

5. R. Rainwater, M. Spira, D. Zeppenfeld, hep-ph/0203187

6. Z. Kunst et al., Z. Phys. C 74, 479 (1997)

6. S. Dawson, Nuc. Phys., B249, 42-60, (1985)

8. U. Baur, and E.W.N. Glover, Nucl. Phys., B347, 12-66 (1990)

9. F. Gianotti et al., CERN-TH/2002-078 (2002)

10. J. Bagger et al., Nucl. Phys. B399, 364 (1993)

11. J. Bagger et al., Phys. Rev. D, 49, 3, 1246 (1994)

12. J. Bagger et al., Phys. Rev. D, 52, 7, 3878 (1995)

13. H-J He et al., ArXiv:hep-ph/0211229, Dec. 20, 2002

14. B. Zhange et al., arXiv:hep-ph/0303048, Apr. 26, 2003